# Comparative study of the Pros and Cons of Programming languages Java, Scala, C++, Haskell, VB .NET, AspectJ, Perl, Ruby, PHP & Scheme

Revision 1.0

Venkatreddy Dwarampudi
Concordia University
Montreal, Quebec, Canada
v_dwaram@yahoo.in

Shahbaz Singh Dhillon
Concordia University,
Montreal, Quebec, Canada
dhillonshahbaz1@gmail.com

Jivitesh Shah
Concordia University,
Montreal Quebec, Canada
ji_shah@cse.concordia.ca

Nikhil Joseph Sebastian
Concordia University,
Montreal Quebec, Canada
n_sebas@encs.concordia.ca

Nitin Satyanarayan Kanigicharla
Concordia University,
Montreal, Quebec, Canada
kanigicharla.nitin@gmail.com

**Abstract**

With the advent of numerous languages it is difficult to realize the edge of one language in a particular scope over another one. We are making an effort, realizing these few issues and comparing some main stream languages like Java, Scala, C++, Haskell, VB .NET, AspectJ, Perl, Ruby, PHP and Scheme keeping in mind some core issues in program development.

## 1. Introduction

### 1.1 Related Work

We were influenced by the study and related work regarding various programming languages and the general discussions, literature provided [1,2,3,4,5,6,7], which lead to the consolidated and precise comparative study of various popular and widely used programming languages by us.

## 1.2 Overview

We give a brief introduction to the languages Java, Scala, Haskell, VB .NET, C++, AspectJ, Perl, Ruby, PHP and Scheme in the section from 1.3 to 1.12. In the next section we provide a consolidated analysis of a pair of languages considering a selected few criteria. The section following this has a concise table of the analysis made in the previous section based on the criterias. Section 4 explains which of the languages are better considering the constraints and purpose of the problem. All the analysis and inferences are followed by supporting references. The later sections consist of Acknowledgement, Abbreviations and some compilable code snippets.

## 1.3 Java

A high-level programming language developed by Sun Microsystems. Java was originally called OAK, and was designed for handheld devices and set-top boxes. Oak was unsuccessful so in 1995 Sun changed the name to Java and modified the language to take advantage of the burgeoning World Wide Web. Java is an object-oriented language similar to C++, but simplified to eliminate language features that cause common programming errors. Java source code files (files with a .java extension) are compiled into a format called bytecode (files with a .class extension), which can then be executed by a Java interpreter. Compiled Java code can run on most computers because Java interpreters and runtime environments, known as Java Virtual Machines (VMs), exist for most operating systems, including UNIX, the Macintosh OS, and Windows. Bytecode can also be converted directly into machine language instructions by a justin-time compiler (JIT). Java is a general purpose programming language with a number of features that make the language well suited for use on the World Wide Web. Small Java applications are called Java applets and can be downloaded from a Web server and run on your computer by a Java-compatible Web browser, such as Netscape Navigator or Microsoft Internet Explorer.

## 1.4 Scala

*Scala* is a language that addresses the major needs of the modern developer. It is a statically typed, mixed-paradigm, JVM language with a succinct, elegant, and flexible syntax, a sophisticated type system, and idioms that promote scalability from small, interpreted scripts to large, sophisticated applications.[8] Scala is used because of it advantages like conciseness, elegance and type-safety.Object-oriented and functional programming are both integrated well thus helping java developers also to be productive with Scala. Talking about conciseness, it actually reduces a code by two to three times as compared to it counter-part Java. So what different about Scala? It helps to integrate functional and Object oriented programming without any restrictions unlike Java. The second--and current--step is Scala, which took some of the ideas of Funnel and put them into a more pragmatic language with special focus on interoperability with standard platforms. Scala and Java are related only by the fact that they are compiled to a bytecode and use the JVM. Scala is completely interoperable with Java. Scala translates to Java bytecodes, and the efficiency of its compiled programs usually equivalent to Java. A .NET version of Scala is also available. A first public release was done in of 2003. [9]

## 1.5 C++

**C++** is a statically typed, free-form, multi-paradigm, compiled, general-purpose programming language. It is regarded as a "middle-level" language, as it comprises a combination of both high-level and low-level language features. It was developed by Bjarne Stroustrup starting in 1979 at Bell Labs as an enhancement to the C programming language and originally named *C with Classes*. It was renamed *C++* in 1983.

## 1.6 Haskell

Haskell is a very high-level language which provides a unique "bird's-eye view" on many programming problems. Like other modern functional languages, Haskell derives its power from higher-order functions, parametric polymorphism and pattern-matching over algebraic data types. Haskell also offers the security of strong, static typing and the flexibility of polymorphism, a combination which helps forestall programming errors without a heavy syntactic overhead. In fact, Haskell's sparse syntax has been specifically designed to be reminiscent of mathematical notation and thus will be familiar to most people. Finally, Haskell features a "pure" mathematical semantics which supports equational reasoning, thus simplifying and streamlining the process of program development. [10]

## 1.7 VB .NET

Visual basic .Net is an object oriented paradigm. Visual Basic .NET is Microsoft's Visual Basic on their .NET framework. Visual Basic .NET is Microsoft's Visual Basic on their .NET framework. Any programmer can develop applications quickly with Visual Basic. It is a very user-friendly language. All you have to do is arrange components using visual tools and then write code for the components. Most programmers of Visual Basic use Visual Studio for their development needs. Moving forward, Microsoft's .NET framework is composed of pre-programmed code that users can access anytime. This pre-programmed code is referred to as the class library. The programs in the class library can be combined or modified in order to suit the needs of programmers. Programs in .NET run on the CLR or the Common Language Runtime environment. Regardless of computer, as long as this environment is present, programs developed in a .NET language will run.

## 1.8 AspectJ

AspectJ is an aspect-oriented programming paradigm .AspectJ is the realization that there are issues or concerns that are not well captured by traditional programming methodologies. Consider the problem of enforcing a security policy in some application. By its nature, security cuts across many of the natural units of modularity of the application. Moreover, the security policy must be uniformly applied to any additions as the application evolves. And the security policy that is being applied might itself evolve. Capturing concerns like a security policy in a disciplined way is difficult and error-prone in a traditional programming language.

## 1.9 Perl

Perl is a high level, general purpose, dynamic programming language. It was designed to be easy for humans, rather than, easy for computers to understand. The syntax of the language is lot more like human language than strict structures. It is very portable as is available for huge variety of operating systems and computers. [11] Perl became popular for two major reasons: First, most of what is being done on the Web happens with text, and is best done with a language that's designed for text processing. More importantly, Perl was appreciably better than the alternatives at the time when people needed something to use. [12]

## 1.10 Ruby

Ruby is a dynamic, reflective, general purpose object-oriented programming language that combines syntax inspired by Perl`s pragmatism with Smalltalk`s conceptual elegance, Python`s ease of learning like features. Ruby supports multiple programming paradigms, including functional, object oriented, imperative and reflective. It also has a dynamic type system and automatic memory management.

**Ruby on Rails**

Rails is an open source frame work for developing database – backed web application. We can develop a web application that is ten times faster than typical java frame work, without compromising in quality of your application because of Ruby programming language many things that are too simple in Ruby are not even possible in most other languages .The guiding principles are less software and convention over configuration. Less Software means fewer lines of code, to implement, code is small means for faster development and fewer bugs makes code easier to understand, maintain and supports enhancements. Convention over configuration means to verbose XML configuration files, there aren't any in Rails. Instead of configuration they are few simple programming conventions that allow it to figure out everything through reflection and discovery. Your application code and your running database already contain everything that Rails need to know.

## 1.11 PHP

PHP: Hypertext Preprocessor is one of the most common web/general purpose scripting languages that produce dynamic web pages. HTML code allows PHP to be embedded within it and interpreted by a web server with a PHP processor module. As a general-purpose programming language, PHP code is processed by an interpreter application in command-line mode performing desired operating system operations and producing program output on its standard output channel. Most modern web servers have PHP processors and most of the operating system have a standalone interpreter. PHP was originally created by Rasmus Lerdorf in 1995. The main implementation of PHP is now produced by the PHP Group and serves as the *de facto* standard for PHP as there is no formal specification. PHP is a free software.[13]

## 1.12 Scheme

Scheme is a statically scoped and properly tail-recursive dialect of the Lisp programming language invented by Guy Lewis Steele Jr. and Gerald Jay Sussman. It was designed to have an exceptionally clear and simple semantics and few different ways to form expressions. A wide variety of programming paradigms, including imperative, functional, and message passing styles, find convenient expression in Scheme. Scheme was one of the first programming languages to incorporate first class procedures as in the lambda calculus, thereby proving the usefulness of static scope rules and block structure in a dynamically typed language. Scheme was the first major dialect of Lisp to distinguish procedures from lambda expressions and symbols, to use a single lexical environment for all variables, and to evaluate the operator position of a procedure call in the same way as an operand position. By relying entirely on procedure calls to express iteration, Scheme emphasized the fact that tail recursive procedure calls are essentially goto's that pass arguments. Scheme was the first widely used programming language to embrace first class escape procedures, from which all previously known sequential control structures can be synthesized. More recently, building upon the design of generic arithmetic in Common Lisp,

Scheme introduced the concept of exact and inexact numbers. Scheme is also the first programming language to support hygienic macros, which permit the syntax of a block-structured language to be extended reliably.

## 2. Analysis

### ✓ Java vs Scheme

***Default more secure programming practices***

- Java provides more secure programming practices as compared to Scheme.

***Web applications development***

- Developing web applications in Scheme compares favourably to developing with Java language.
- We have to spend much effort developing libraries and fixing errors that would not have been an issue with a more mature platform, but we can use a range of language features not available elsewhere. After an initial one-off 'startup' cost this tradeoff works.

***Web services design and composition***

- Web services depend on the ability of enterprises using different computing platforms to communicate with each other. This requirement makes the Java platform, which makes code portable, the natural choice for developing Web services. This choice is even more attractive as the new Java APIs for XML become available, making it easier and easier to use XML from the Java programming language.In addition to data portability and code portability, Web services need to be scalable, secure,and efficient, especially as they grow. The Java 2 Platform, Enterprise Edition (J2EE), is specifically designed to fill just such needs. It facilitates the really hard part of developing Web services, which is programming the infrastructure, or "plumbing." This infrastructure includes features such as security, distributed transaction management, and connection pool management, all of which are essential for industrial strength Web services. And because components are reusable, development time is substantially reduced.

***OO-based abstraction***

- The Scheme code is a bit verbose, with obvious redundancy between the internal function definitions and the dispatch methods. This could be reduced fairly easily with some simple macros to define methods as used in Java.

***Reflection***

- The reflection mechanism in Scheme is not as powerful as Java. Java supplies a rich set of operations for using metadata and just a few important intercession capabilities. In addition, Java avoids many complications by not allowing direct metaobject modification. These features give reflection the power to make your software flexible. The software marketplace is increasing its demand for flexibility. Knowing how to produce flexible code increases your value in the marketplace. Java is so well crafted and its reflection API so carefully constrained that security is controlled simply.

***Aspect Orientation***

- This three step process of mimicking the operators of AspectJ but implementing them in the context of the Scheme programming language requires significant amount of design effort. On the other hand AspectJ is a complete implementation of an AOP language for Java. It consists of weavers that take various forms such as a compiler and a linker. A weaver produces byte code that conforms to the Java byte-code specification, allowing any compliant JVM to execute those class files. It is easy to learn and use. To simplify building and debugging applications, the language implementation also offers support for IDEs. AspectJ enables Java developers to better manage the problems in large program systems and to reap the benefits of modularity.

***Functional programming***

- Java has no functions; however, using interfaces and inner classes it is possible to mimic some but not all the features of functional programming. But Scheme is primarily a functional programming language. Unlike most other functional languages, Scheme supports multiple coding paradigms, and functional programming is a subset of its capabilities. However, its functional capabilities are complete -it's not lacking anything that a functional language needs to be considered functional. It also has many useful built-in functions (arithmetic, list operations, etc). The Scheme functional programming style meshes well with both multi-threading and transactional based systems. If you can get the logic correct, functional programming in Scheme requires orders of magnitude less code. That means fewer points of failure, less code to test, and a more productive programming life.

***Declarative programming***

- You can embed some of the declarative programming features in Java using libraries like JSetL and JSolver. But Scheme is a general purpose language which offers a powerful and flexible declarative programming model because it is in clear correspondence to mathematical logic and lacks side effects. The programs in Scheme are concise; this makes it easy even for nonprogrammers to obtain solutions.

***Batch scripting***

- Scheme Shell is a system with several faces. From one perspective, it is not much more than a system-call library and a few macros. Yet, there is power in this minimalist description—it points up the utility of embedding systems in languages such as Scheme. A Scheme shell wins because it is broad-spectrum. A functional language is an excellent tool for systems programming.

***UI prototype design***

The Scheme code is a better choice because of the following reasons:

- There’s less preamble (#lang scheme/gui vs multiple imports).
- The syntax is consistent with what has come before in the student’s experience. Hence it is not very difficult for a stakeholder to understand.
- Less number of lines of code.
- There aren’t any mystery calls (frame.pack()?).
- new is analogous to make-, which students have seen many times before.
- Callback functions follow naturally from previous work; event handlers are confusing.

## ✓ **Scala** vs. **PHP**

***Default more secure programming practices***

- Traditionally PHP is weakly typed vs. Scala which is Strongly typed preventing unexpected invalid input behavior.[15]

- Input validation and input filtering not directly possible in traditional PHP application.[15]
- Scala is a type safe programming language.[14]

***Web applications development***

- PHP integrates HTML code effortlessly; this helps web servers to process the web pages before they are actually displayed on the web browser. Scala can possess this feature but not effortlessly, since it is not full-fledged web development programming language.[16]
- PHP is dynamic web development language whereas in Scala need frameworks to perform web development.[16]
- PHP since is versatile server-side web development program and is supported on most of the web servers and runs on all operating systems. Scala needs the JVM installed and other add-on to support web development.
- Since PHP is faster it helps in faster page loading[16]
- There is huge repository of documentation available for web development in PHP[16]

***Web services design and composition***

- PHP is reasonably high performing implementations while providing users with high software productivity.PHP is a viable option for publishing SOAP/WS-* based web services in addition to the currently popular REST-style web services.[17]
- Scala's have an edge over PHP in categories like Security, maintainability while PHP has better performance due to dynamic compilation of code. PHP has high flexibility and excellent documentation providing a broader view about the web services it can offer.
- PHP has a wider support for mostly all web service designs while Scala being a relatively new language is trying to make a mark in the web development and service world

***OO-based abstraction***

- Lines of Code: For Scala, the lines of codes is less than that of PHP
- Readability: The readability for PHP is better but the preciseness and conciseness is obtained in Scala.
- Program load time: For Scala there is a new and unique feature provided by the compiler called the FSC (Fast Scala compiler) [18] which does not compile the repeated code and does the compilation in almost no latency, unlike PHP which has to be compiled every time it is executed.
- Instruction Path Length: In scala, this feature is highlighted the most with the instruction path length being very less as compared to PHP just like all other object oriented programming languages.

***Reflection***

- Using Reflective API in PHP is expensive and dents the performance. Scala implements Reflection with some memory foot print which can be dealt with manually.[19]
- Scala allow implementation of Reflection API sharing the same interface using global factory.
- Reflection in PHP is easier to implement than in Scala.

***Aspect-orientation***

- A major difference and a strong point of Scala over PHP is for Aspect oriented programming in Scala, it has a complete Type Safety which the later lacks.

- Earlier versions of PHP were inadequate to satisfy the core reasons of implementing Aspect Oriented programming. Later versions have tried to do the same with some performance penalties.

***Functional programming***

- Scala implements functional programming effortlessly. It is a built-in feature of the Scala language unlike PHP which has it as an extension, besides it's functional calls are verbose.
- PHP lacks certain primitive functional programming features

***Declarative programming***

- Scala Declarative programming has a structure. PHP has the flexibility of creating its own annotations
- Scala can handle declarative programming to a higher extent that declarative programming implemented in PHP since it is equipped with a interpreter

***Batch scripting***

- Scala Has access to the COMPLETE JDK and can put other Java/Scala-Libraries into my Classpath to use in a script[20]
- Scala can do all file operations on a high and object oriented level
- With Scala , it is Small and readable scripts.
- Scala can also include in future Wrapper to create a GUI for the script
- PHP provides very limited security and scripting options as compared to the vast Scala language functionality.[21]

***UI prototype design***

- In Scala, its "everything is an object" philosophy makes it possible to inherit the main method of a GUI application. [20]. PHP does not apply this concept entirely. PHP passes parameter by value and reference and not always objects.
- In Scala, the method can be hidden from user applications, including the boilerplate code for setting things up that comes with it. PHP provides very less security and hiding of method from user application but it can be performed by explicitly mentioning the expose_php value in php.ini which reduces information available to users.[22]
- Scala's first-class functions and pattern matching make it possible to formulate event handling as the reactions component property, which greatly simplifies life for the application developer.[20]

## ✓ **VB .NET** vs. **AspectJ**

***Default more secure programming practices***

- VB.NET supports exception handling programmatically and implicitly.
- Some errors in VB.NET cannot handle implicitly but handled programmatically.
- .NET Framework by default provides security such as internal security, data security and external security.
- AspectJ supports exception handling by extracting error handling code from normal code.[23]
- AspectJ can plug and Unplug error handling code automatically.

- AspectJ limited supported of security but enable programmer to write secure code. Example for default secure entity's are returns and returned pointcuts.

***Web Application development***

- VB.NET support web applications.
- VB.NET support dynamically by accessing database.
- AspecJ support web applications such as JSP and Spring Framework but it cannot do by itself.
- AspectJ increases modularity in web application development and security.

***Web services design & Composition***

- VB.NET can implement by default using .NET Remoting.
- VB.NET also support different web services such as HTTP, WSDL, SOAP and UDDI.[24]
- AspectJ can deploy an Axis web service.
- AspectJ provides dynamic switching between different technologies.

***OO-based Abstraction***

- VB.NET was an object oriented programming language.
- VB.NET can implement Abstraction.
- AspectJ can implement Abstraction but violates object oriented feature encapsulation by accessing private members outside the class.
- Readability of VB.NET programming is good for basic programmer as compared with AspectJ
- Lines of code in AspectJ are more.

***Reflection***

- AspectJ can implement static and dynamic behavior of reflective programming but supports dynamic behavior partially.
- VB.NET completely supports dynamic and static behavior of reflective programming.
- VB.NET supports dynamic behavior can save memory but slows down speed as compared with AspectJ.

***Aspect Oriented Programming***

- AspectJ was an implementation of aspect oriented programming.
- AspectJ can implement both static and dynamic weaving.
- VB.NET can implement aspect oriented programming using AOP Engine [25].
- VB.NET can implement only dynamic weaving only.

***Functional Programming***

- VB.NET supports functional programming partially because can implement lambda calculus but not lazy evaluation.
- AspectJ doesn`t support functional programming.

***Declarative Programming***

- VB.NET can support declarative programming by depending on different technologies such as XAML,LINQ [26]

- AspectJ support declarative programming by annotations[27]

***Batch Scripting***

- VB.net uses "System.Diagnostics.Process" package to call bat file.
- Macros can be written by using VB.NET
- AspectJ doesn't support batch scripting itself but it can use java libraries to execute.

***UI prototype design***

- VB.NET by default support UI prototype
- VB.NET gives rich look to UI and provides drag and drop facility to the programmer.
- AspectJ increases performance and enable security to the UI.

### ✓ **C++** vs. **Ruby**

***Default more secure programming practices***

- Ruby has better advantages like garbage collection, no pointers, indicates buffer overflow.
- Integer.MAX_value+1 equals Integer. Min_Value in C++, But in ruby it expands the integer as needed.

***Web applications development***

- C++ is very high performance language, but writing scalable, threaded infrastructure code can be very complicated and time consuming but it is done well in ruby.[29]. Since C++ is a compile time language it is difficult to map it with dynamic systems.[29].
- Querying and string manipulations are main causes in web development are good to do in dynamic language than in C++.[28]

***Web services design and composition***

Advantages of WSO2 WSF/C++.[31]

- Convenient and speedy development with built-in tools
- Complete WS-* stack for C++
- Lightweight
- Support for protocols
- Integration with existing environment
- Handle REST and SOAP message formats.[31].

Advantages of WSO2 WSF Ruby.[30]

- Better Performance
- Easy to use Object Oriented API
- Extensible
- Wider Support(SOAP 1.1/SOAP 1.2/SOAP MTOM)
- Built by Web Services Experts

- Industrial Strength

***OO-based abstraction***

- In both C++ and Ruby, public methods can be called by anyone, and protected methods can be called only by objects of the same class or objects that inherit from the defining class. The semantics of private differ between C++ and Ruby, however. In C++, methods are private to the class, while in Ruby they are private to the instance. In other words, you can never explicitly specify the receiver for a private method call in Ruby.[32]
- Ruby, unlike C++, does not support multiple inheritance but does support *mixins*. You can't create an instance of Module; you can only include it into class definitions. In this case, Module Comparable defines the comparison operators (<, >, ==) in terms of the <=> operator. So by defining <=> and including Module Comparable, you get the other comparison operators for free.
- In C++, you sometimes rely on inheritance combined with virtual functions to enable polymorphism. A pointer x of type T * can point to an object of type T or any object with a type below T in the class hierarchy. A virtual method invoked through x is resolved by walking up the class hierarchy, starting from the type of the object pointed to by x.
- Ruby on the other hand uses duck typing—*if something looks like a duck, swims like a duck, and quacks like a duck, then it's a duck*.
- For Ruby, it doesn't matter what type x is. If the object x has a method print_hello, the code will work. So unlike C++, which would require the objects to inherit from a common base type, you can pass objects of unrelated types to my_method, as long as they all implement print_hello.[32]

***Reflection***

There are two types of reflection

- Inspection by iterating over members of a type, enumerating its method and so on. This type is not possible in C++ but possible in Ruby , as its meta programming tool kit supports all types of meta data. For achieving this in C++ there is no direct way but using a meta compiler like qt meta object compiler which translates code by adding additional meta information and even high level inspection is not possible.[33].
- Inspection by checking whether a class type (class,struct,unions) has a method or nested type, is derived from another particular type. This is possible in C++ using template tricks like use of boost::type_traits for many things. For checking whether a certain nested type exists use plain SFINAE. [33]. Ruby supports even this type of reflection.

***Aspect Orientation***

- AspectR is a little different, as it doesn't introduce any new syntax into the language and implements AOP using it's standard OO programming unlike AspectC++. This is possible due to Ruby's dynamic nature.
- AspectC++ uses preprocessing to weave the advice, this means a pass through the code is required before passing it to the standard C++ toolchain. AspectR does all it's work at runtime and doesn't modify the language in any way. In Ruby there isn't any difference between compile time and run time so this makes sense. AspectR's approach clearly adds runtime overhead but might offer greater flexibility.

- The example shows how to use AOP to weave advice that is run before and after a method call. There are three methods in our test class, TestClass: method1, method2 and method3. All methods take a string as argument and return another. In method1 AOP is used to weave advice that runs before the method and prints the argument passed to it. In method2 advice is weaved to run after the method, printing the value that was returned by the method. In method 3 advice is added around the method, changing it's arguments and return values.[34].

***Functional programming***

- Both the languages do not support pure functional programming. But it can be achieved in an extreme extent in C++ using FC++, but when comparing only C++ and Ruby, Ruby is the better functional language than C++.Both of the languages do not force to write pure functions but certainly they help to do. But we can expect more help from Ruby as they are mentioned above.

***Declarative programming***

- As C++ is an imperative programming and it will not support declarative style by default. But efforts are made to implement declarative programming in C++.These are some of the ways to implement declarative programming in C++.[31]Pure logic programming is entirely declarative in nature. The primitives used to support logic programming are provided by the **caster**, an open source C++ library work. Castor is small , pure C++ header library. It seamless integrates logic programming paradigms supported by C++ such as object oriented programming, imperative, generic etc. Blending logic programming techniques enables new designs and programming techniques that are not available to purely logic programming approaches or pure object oriented approaches, but only feasible with multi paradigm blend. A key goal of caster is to smooth and efficient blend into rest of C++ with minimal syntactic over head.[35].
- COP(C++ or Prolog), some efforts made by Charles- Ant oine Brunet, Ruben Gonzalez Rubio to integrate C++ and Prolog for declarative type of programming possible in C++. Integration of C++ and Prolog into one named as COP(C++ or Prolog). The approach has been made to add some features in C++ to achieve Prolog goals. This approach is clumsy because a programmer must make some extra manipulations. The great disadvantages are that the bugs in programs can be hard to discover and the code depends on the platforms. The proposal is to merge two different languages into one language keeping the characteristics of the two. Goals of proposed language are

  -not to change syntax and semantics of both languages.
  -to link between new concepts that link between C++ and Prolog
  -to handle errors
  -to reduce over head

***Batch scripting***

- Undoubtedly Ruby holds upper hand than C++ in scripting. Ruby is inbuilt with number of files for scripting purposes i.e to execute external programs, command line options and arguments, shell library, accessing environment variables. C++ is preferred for the criteria performance is the factor but loading some library files is suppressing it.

***UI prototype design***

- We can use the frame works provided by the Qt/GTK+ etc for graphical user interface design for platform independent applications either in C++ or in Ruby. But the complexity of syntax in C++ will generally make it difficult to use. We even can build a window without these libraries but it is a difficult work to be done. As the example mentioned above will give the same output for but one can observe the difference in complexity and amount of issues to be handled while creating a simple window, all these will be done by default. So, Ruby is preferred for user interface design where complexity can be minimized but in the applications that are to be differently behaved should prefer C++.

## ✓ **Haskell** vs. **Perl**

***Default more secure programming practices***

- Garbage collection
  Both Haskell and Perl have their own Garbage collection mechanism. However, the garbage collection mechanism in Perl might break down when one has a circle of reference values. On the other hand, garbage collection for Haskell has no such drawbacks and has a better garbage collection mechanism than Perl.
- Pointers
  Both Haskell and Perl does not have pointers, this means that a programmer is not allowed to play with pointers and avoiding memory mismanagement.
- Type System
  As mentioned above, Haskell has a strong Type System security. Whereas, in Perl one cannot safely code without numerous checks. A drawback of Perl's Type System.

***Web applications development***

- As mentioned above both Haskell and Perl are a good choice for web development. The code sample shows that there is not much of a difference in the code size as well. However, when it comes to web development for large applications, Perl is a better choice than Haskell.

***Web services design and composition***

- Both Haskell and Perl can be used for web service. However, Perl is a better choice as it provides more services than Haskell, such as WSDL and UDDI. Also, as mentioned above, full support to WSDL to Haskell is still under development.

***OO-based abstraction***

- Object-Oriented Concepts are not supported by Haskell. It requires an extension OOHaskell to implement these concepts. However, in Perl, there is a straight forward implementation of these concepts.

***Reflection***

- Apparently, Perl has a better reflection mechanism than Haskell. Haskell do have libraries for dynamics, but they still do not support complete reflection.Perl on the other hand, has good reflection mechanisms.

***Aspect-orientation***

- Both Haskell and Perl support Aspect Orientation. However, as mentioned above, Perl uses an easier AO implementation mechanism than Haskell. This is because Haskell has to use an extension called AOP Haskell whereas in Perl there is an in build package Aspect. Hence, Perl is a better choice. Also, from the code sample above, we can notice that Perl has an easier AO implementation than Haskell.

***Functional programming***

- Though Perl has some support for Functional style of Programming with the help of References and Closures, Haskell is built as a Functional Language and is better for functional style of programming than Perl.

***Declarative programming***

- Perl has many limitations when it comes to Declarative / Logic programming as mentioned above. However, with Haskell, one can use Monards to implement Declarative/ Logic programming. Therefore, Haskell is well suited for Declarative style of Programming.

***Batch scripting***

- Shell Scripting is possible with Haskell using HSH that was released in 2007. However, Perl is a Scripting language and is more suited for shell scripts than Haskell.

***UI prototype design***

- As mentioned, both Haskell and Perl have rich set of libraries to build a GUI application. I have just shown the description of one library each for Perl and Haskell, however, there are many more that each programming languages support.

## 3. Consolidated Analysis and Synthesis of the Results

| Criteria/PL | Java | Scala | C++ | Haskell | VB .NET |
|---|---|---|---|---|---|
| **Default more secure programming practices** | Good secure programming features with GC, no pointers, packages and threads. | Good Default Security with features like GC, Exception handling & works on JVM so uses its Security Manager | not a secure programming language, buffer overflow is not detected. | Good secure programming features with a GC, no pointers and good type system. | Built-in secure features provided by .net and programming itself can implement secure features |
| Web Applications | Quite popular for web applications. Abundant libraries and servlets serve this cause. | Can develop flexible, highly scalable,secure applications with help of web development frameworks | Used for standalone applications, difficult to create by default. | Can develop we applications with rich set libraries. | Vb.net supports web applications |
| **Web Services Design and Composition** | Good for web services because of portability and large number of APIs for XML available. | RESTful services provided with help of frameworks. Provision of other services still under construction. XML processing simple | Supports REST,XML, WSO2 frame work | Provides services like SOAP and REST but is still immature in terms of WSDL and UDDI. | Vb.net can implement web services such as HTTP,SOAP, XML , WSDL , UDDI and .Net remoting service can implement it self |
| Object-Oriented based Abstraction | Primarily an object oriented language with powerful features. | Supports 2 types of abstraction .Alternative to functional abstraction. Mainly used for modeling families that vary covariantly | Supports Object Oriented principles but not as a default. | Object-Oriented Concepts are not supported by Haskell. It requires an extension called OOHaskell. | Vb.net is an object oriented language .It supports OO abstraction. |
| **Reflection** | Powerful reflection mechanism. Supplies a rich set of operations for using metadata and avoids complications. | Its a subsystem, Reflection API..Limited scope.Modular, hence reduce foot-print & be efficient | Limited reflection capabilities. | Haskell do have libraries for dynamics, but they still do not support complete reflection. | Vb.net supports reflection using built-in called “system.reflection” |
| Aspect-Oriented Programming | AspectJ, an extension to Java treats AOP concepts as first-class elements of the language. | Provides 2 different types.Mainly, Mixin composition stacks | With static type of language it is difficult , AspectC++ supports it | Does not directly support. Has an extension called AOP Haskell. | Aop Engine in .NET to implement AOP programming but it supports only at run time |
| **Functional Programming** | No functionsInstead, using interfaces & inner classes it is fairly easy to mimic some features of FP. | Powerful Support and well suited. Light-weight syntax. Supports High-order, nested functions, and currying | Doesn’t support to fuller extent but can be done using FC++ | This is a functional programming langage. | Vb.net is not a pure functional programming but it supports Lambda calculus |
| Declarative Programming | Libraries like JSetL and JSolver offer a number of facilities to support DP. | Uses a Prolog interpreter called ScalaLogic. Emphasises on Simplicity and not performance | By default not possible but merging prolog is an alternative. | Haskell, one can use Monards to implement Declarative/ Logic programming. | Dosen’t implement declarative programming by itself |
| **Batch Scripting** | Easy; involves the use of two Java classes, the Runtime class and the Process class. | Supports Batch/Bash/Perl scripting. Used as real scripting language | Including libraries allows to do so. But decreases performance. | Shell Scripting is possible with Haskell using HSH. | Vb.net supports batch scripting and macros |
| UI prototype design | Rich set of libraries for UI applications but the code is verbose and can be mysterious for stakeholders. | Supports UI with basis on Java swing framework but hides much of its complexity | Difficult to implement by default but supports some libraries. | Has rich set of libraries for GUI applications. | vb.net supports rich UI interfaces and IDE give good support to programmer. |

| Criteria/PL | AspectJ | PHP | Ruby | Perl | Scheme |
|---|---|---|---|---|---|
| **Default more secure programming practices** | It capture the returned values of methods in both the execution and method invocation. It implements execution handling itself. | Insecure, weakly typed language. No input validation, etc Security provided with help of few frameworks. currently. | secure, dynamic and has GC. No pointers, Exception handling. | Has no proper GC mechanism, has no pointers but references and no type system hence, one cannot safely code. | Less pros and more cons. Problems with types, records, threads, etc. |
| Web Applications | It supports web applications by proving secure features during transactions | Excellent features. Dynamic easy less code required. Can be embedded into HTML | Using Rails , supports for fast and secure applications | Has excellent features for web application development. | Powerful features for web development but fewer libraries than mainstream. |
| **Web Services Design and Composition** | AspectJ supports web services such as AXIS but doesn't support all features by itself. | Good choice for web services Provides SOAP support inbuilt in PHP5.Supports REST,WS-* | Supports many frame works as it is a flexible language. | Is indeed a good choice for this as it has almost all the features. | PLT Scheme has good web services support. |
| Object-Oriented based Abstraction | It supports OO Abstraction | Not ideal for OO-based abstraction, but has a clear and simple implementation | By default it is object oriented language and supports abstraction. | Has a straight forward implementation. | Closures can be used to capture object state. But the code is verbose with obvious redundancy. |
| **Reflection** | It can implement Reflection using " "org.aspectj.lang.reflect.*" | Implemented using standard included Reflection API | Supports reflection programming. | has good reflection mechanisms. | Homoiconic language; can meta program during runtime. |
| Aspect-Oriented Programming | AspectJ itself is an AOP programming. It can support at static and dynamic time. | Supported using Libraries. Aspects are statically weaved. Dynamics weaving is possible using extensions | Supports AOP and can be done with Aspect Ruby | Has a straight implementation with a built in package called Aspect. | Continuation marks and language-defining macros help in implementing AOP in Scheme. |
| **Functional Programming** | AspectJ implement a profiler that records statistics concerning the number of calls to each method**.** | Supports but its functional calls are verbose. Few FP primitives introduced to improve | Supports but doesn't force to do it. | Some support with the help of References and Closures. | Primarily a functional programming language. |
| Declarative Programming | It supports declarative programming using Annotations | Do not support by default. Uses lesser known technique of Annotations and has considerable limitations | By default not possible merging prolog is an alternative | Perl has many limitations when it comes to Declarative / Logic programming. | Capable of doing declarative programming. Schelog and Kanren serve this cause. |
| **Batch Scripting** | It doesn't support itself | Supports but with limited advantage. Must be compiled as a CGI binary | Supports scripting . | It is a scripting language and is well suited for this. | Some Schemes allow to define Unix-style scripts containing Scheme code |
| UI prototype design | It increases UI performance such as fast response of UI | Basic version does not support. Implemented using supporting frameworks | Very easy to implement. | Has rich set of libraries for GUI applications. | Good for UI applications. Syntax is consistent and easy for stakeholders. |

## 3.1 Criteria 1: Default Secure Programming Language

After careful study regarding the language mentioned, we came to the conclusion that Default security was best provided by Java, Scala, VB .NET and Haskell. All the three languages have an automated garbage collector. Scala provides implicit exception handling and since it run on the JVM, it work in collaboration with the Java security manager, this is one of the many strong points Scala has over Java's default security features. Threads(Actors in Scala), packages, Buffer overflow exceptions are some other default security provided by Java, VB .NET and Scala. Haskell has no pointers and has a good type system. As for other languages like Scheme, Perl, Ruby provides security but lacks few more secure features .

## 3.2 Criteria 2: Web Applications Development

With Respect to Web application development PHP and Ruby(with ROR) are the best preferred languages. Ruby is most preferred in cases where security, performance, readability and flexibility is concerned. PHP is very widely used due to its simplicity, huge repository of documentations and light weight syntax, but by default it lacks security. Java, Scala(eg. Twitter, LinkedIN) and Ruby is preferred for cases where the Web application demands security and robustness without compensating largely on the performance of the application. C++ develops standalone application which can be used where performance is the main criteria.

## 3.3 Criteria 3: Web Services Design and Composition

Turning our attention to Web Services Design and Composition, PHP and Perl are the preferred choices. PHP supports an extensive set of Web services and still adding. SOAP is an inbuilt web services in PHP version 5..As for Perl, Web services are provided using wide rich set of libraries. Java, Ruby and C++ provide web services and excellent XML support with libraries, API's and many frameworks. As for Scala and Haskell Web services are provided but are immature.

## 3.4 Criteria 4: Object Oriented based abstraction

Ruby is a pure object oriented language and is best suited for this purpose. Scala, Java, C++ and VB .NET can be considered as substitutes for this purpose. Scala's advantage over Java are based on the compactness, easy scalability and efficiency. Java provides macros that help improve efficiency to a considerable extent. Haskell provides support for OO-based programming in it extension OOHaskell. Scheme is long, verbose and redundant whereas AspectJ has dependencies, hence does not suffice the purpose of OO- based abstraction.

## 3.5 Criteria 5: Reflection

VB.NET completely supports reflective programming such as access to the static and dynamic information but AspectJ supports dynamic crosscutting concern partially.java and Scala very well crafted that it provides excellent security inspite of reflection. Both scheme and Perl support reflective programming using Homo iconic language but they are not considered among the best.PHP dents the performance and uses lot of memory. Ruby uses rich set of libraries for the support of reflective programming. Haskell has limited reflective programming support.

## 3.6 Criteria 6: Aspect Oriented Programming

AspectJ was an implementation of aspect-oriented programming .VB.NET supports AOP using AOP Engine.java, scala .Perl, Ruby and Haskell provides aspect oriented programming using extensions. Scheme requires design effort. C++ has a static nature due to that AOP is difficult.

## 3.7 Criteria 7: Functional Programming

Haskell, Scala and Scheme are implementations of functional programming. Both Ruby and c++ doesn't force you to write in pure functional programming but it supports functional programming and C++ doesn't support higher order functions.VB.NET supports functional programming partially such as lambda calculus but doesn't support lazy evaluation.

## 3.8 Criteria 8: Declarative Programming

Declarative programming in mentioned programming languages cannot be done without frameworks, interpreters or extensions.. Scheme does this using Schelog and Kanren whereas Scala has a prolog interpreter called ScalaLogic, Haskell uses Monards, C++ uses COP, Ruby uses RubyProlog and so on. PHP and Perl are not good options for this kind of programming. Declarative programming in PHP and Perl have a very limited scope.

## 3.9 Criteria 9: Batch Scripting

Perl is a scripting language and best suited for this purpose. It is easy, small and portable. Java, Ruby ,Scala and VB .NET are also good considering its ability to perform automation, macros and shell scripting on different platforms. C++ is mainly performance oriented language as it requires some libraries for scripting and which degrades its performance.

## 3.10 Criteria 10: UI prototype design

Haskell, VB.net, Perl, Ruby has good repository set of libraries for GUI design at a faster way. Both java and Scala use swing framework, GUI design. Scheme is also one of the subtly preferable UI language due to preamble, small syntax consistent and has no mystery calls. PHP also supports comparatively decent UI using mature extensions/frameworks. C++ is a tedious way for creating an UI design but C++ with framework provides better support. AspectJ cannot provide UI design because of it uniqueness to perform aspect oriented programming. For cases where it is required it depends on Java & its libraries.

## 4. Conclusion

After carefully analyzing the languages, we have come to the conclusion that every languages has it's own ups and downs. Every particular language has a purpose but can be extended or revised to accommodate the current needs of programming. Inspite of all this, every language has it own speciality and considerably better programming practices which has made it popular and revolutionized the computing world. In the days to come, languages like Scala are expected to introduce some new and challenging perspective to the programming practices which started it journey from procedural to Object oriented, Declarative, Reflective, functional to a combination of these features. As for now many languages are at the advent of change and introducing new features and competing with each other. Keeping this in mind, we conclude that each language has a specialized task and have to analyzed and gauged for implementation.

### 4.1 Acknowledgements

I would like to acknowledge the contribution of the following entities that made this presentation possible

- Professor Serguei A. Mokhov, for the advise, topics, the help in understanding the concepts, the direction of research for the criteria's and immense support.
- Faculty of Engineering and Computer Science, Concordia University, Montreal, Canada
- Concordia University Libraries for access to the invaluable digital libraries ACM, IEEE, Springer and for the books.
- POD, Yi Ji for introduction into AspectJ[36] and Java Reflection[37]
- Wikipedia, contributors of a wealth of information
- Martin Odersky for innumerable papers, books, suggestions for the Scala Language research.

## Abbreviations

| | |
|---|---|
| VB | Visual Basic |
| PHP | Preprocessor |
| JVM | Java Virtual machine |
| VM | Virtual Machine |
| XML | Extensible Markup Language |
| HTML | HyperText Markup Language |
| API | Application Programming Interface |
| IDE | Integrated Development Environment |
| AOP | Aspect Oriented Programming |
| SOAP | Simple Object Access Protocol |
| REST | Representational State Transfer |
| WSDL | Web Service Definition Language |
| HTTP | HyperText Transfer Protocol |
| UDDI | Universal Description Discovery and Integration |
| GUI/UI | Graphic User Interface/User Interface |
| XAML | Extensible Application Markup Language |
| LINQ | Language-integrated query |
| JDK | Java Development Kit |
| WSO2 | Web Services Oxygen |
| WSF | WorkStation Functions |
| SOAP MTOM | Simple Object Access Protocol Message Transmission Optimization Mechanism |
| HSH | Haskell Shell |
| FP | Functional Programming |

## A. Source code for Object-Oriented Abstraction in a few languages.

- Scala

1. The following code was run on a Scala compiler from http://www.scala-lang.org/downloads.
2. Installing the Java JVM on the system
3. Opening the command prompt by opening “scala.bat” in the unzipped folder of Scala downloaded
4. Run the code provided at the scala command prompt as scala> Objectname.main(null).

✓ **Criteria : Object- Oriented based Abstraction**

```
package polymorph

abstract class Shape(initx:Int, inity:Int){
  var x: Int = initx
```

```
  var y: Int = inity

  def moveTo(newx: Int, newy: Int){
    x = newx
    y = newy
  }
  def rMoveTo(dx: Int, dy: Int){
    moveTo(x + dx, y+ dy)
  }
  def draw() = {}
}

class Rectangle(initx: Int, inity: Int, initwidth: Int, initheight: 
Int) extends Shape(initx, inity){
  var width = initwidth
  var height = initheight
  override def draw() = println("Drawing a Rectangle at:(" + x + "," + 
y + "), width " + width + ", height " + height)
}
class Circle(initx: Int, inity: Int, initradius: Int) extends 
Shape(initx, inity){
  var radius = initradius
  override def draw() = println("Drawing a Circle at:(" + x + "," + y 
+ "), radius " + radius)
}
object Polymorph {
  def main(args : Array[String]) : Unit = {
    // create a collection containing various shape instances
    val shapes = new Rectangle(10, 20, 5, 6) :: new Circle(15, 25, 
8):: Nil
    shapes.foreach(ashape =>{
      ashape.draw()
      ashape.rMoveTo(100,100)
      ashape.draw()
      }
    )
   // access a rectangle specific function
       val rectangle = new Rectangle(0, 0, 15, 15)
       rectangle.width_=(30)
       rectangle.draw()
 }}
```

## PHP

1. Install the WAMP server software available for free online http://www.wampserver.com/download.php
2. At the bottom taskbar, click on start all services
3. Paste the code in a notepad file and save it in the wamp folder /www.
4. Run the code on a Web browser by opening it.

- ✓ **Criteria : Object oriented based abstraction**

**File: Shape.php**

```
<?php
interface Shape {
	public function Draw();
	public function MoveTo($x, $y);
	public function RMoveTo($dx, $dy);
}
?>
```

**File: Rectangle.php**

```
<?php
class Rectangle implements Shape {
	// If a method can be static, declare it static. Speed
improvement is by a factor of 4. (http://vega.rd.no/article/php-
static-method-performance)
	private static $width;
	private static $height;
	private static $x;
	private static $y;
	function __construct ($x, $y, $width, $height) {
		self::$x      = (float) $x;
		self::$y      = (float) $y;
		self::$width  = (float) $width;
		self::$height = (float) $height;
	}
	public function Draw () {
		echo sprintf("Drawing a Rectangle at %d, %d, width %d,
height %d<br>", self::$x, self::$y, self::$width, self::$height);
	}
	public function MoveTo ($x, $y) {
		self::$x = $x;
		self::$y = $y;
	}
	public function RMoveTo ($dx, $dy) {
```

```
                self::$x += $dx;
                self::$y += $dy;
        }
        public static function __set ($var, $val) {
        self::$$var = (float) $val; // note the '$$'
        }

    public static function __get ($var) {
         return  self::$$var; // note the '$$'
    }
}?>
```

**File: Circle.php**

```
<?php
class Circle implements Shape {
        // if a method can be static, declare it static. Speed
improvement is by a factor of 4. (http://vega.rd.no/article/php-
static-method-performance)
        private static $radius;
        private static $x;
        private static $y;
        function __construct($x, $y, $radius) {
                self::$x      = (float) $x;
                self::$y      = (float) $y;
                self::$radius = (float) $radius;
        }
        public function Draw () {
                echo sprintf("Drawing a Circle at %d, %d, radius %d<br>",
self::$x, self::$y, self::$radius);
        }
        public function MoveTo ($x, $y) {
                self::$x = $x;
                self::$y = $y;
        }
        public function RMoveTo ($dx, $dy) {
                self::$x += $dx;
                self::$y += $dy;
        }
        public static function __set ($var, $val) {
        self::$$var = (float) $val; // note the '$$'
    }
    public static function __get ($var) {
```

```
        return  self::$$var; // note the '$$'
    }}
?>
```

**File: main.php**

```
<?php
       // load the Circle and Rectangle classes and Shape interface
       function __autoload ($className) {
              require $className . '.php';
       }
       // set up an array with some shape instances
    $shapes = array(new Rectangle(10, 20, 5, 6), new Circle(15, 25,
8));
       // iterate through the shapes
       for ($i = 0; $i < 2; $i++) {
       $shapes[$i]->Draw();
       $shapes[$i]->RMoveTo(100, 100);
       $shapes[$i]->Draw();
       }
       echo "<br><br>";
       // circle specific functions
       $circle = new Circle(10, 20, 30);
       $circle->Draw();
       $circle->radius = 100; // set new value with __set "magic
method"
       $circle->Draw();
       $circle->MoveTo(200, 200);
       $circle->Draw();
       echo "<br><br>";
       // rectangle specific functions
       $rectangle = new Rectangle(10, 20, 30, 40);
       $rectangle->Draw();
       $rectangle->width  = 100; // set new value with __set "magic
method"
       $rectangle->height = 195; // set new value with __set "magic
method"
       $rectangle->Draw();
       $rectangle->RMoveTo(10, 10);
       $rectangle->Draw();
?>
```

## AspectJ

- Download “ajdt_2.0.2_for_eclipse_3.5.zip” from [http://download.eclipse.org/tools/ajdt/35/update/ajdt_2.0.2_for_eclipse_3.5.zip](http://download.eclipse.org/tools/ajdt/35/update/ajdt_2.0.2_for_eclipse_3.5.zip) then install it.
- Add “aspectjrt.jar” and “C:\\aspectjhome\\bin to class path
- Compile as ajc *.java
- Execute as java main function class name

✓ **Criteria : Object oriented based abstraction**

```
package abstraction;
public abstract aspect Parent
{
int i=0;
pointcut greeting() : execution(* Main.fun(..));
}
aspect BigWorld extends Parent
{
 after(): greeting()
{
System.out.println("I'm in child class");
 System.out.println("base class member i values is"+i);
}
}
package abstraction;
public class Main
{
public static void main(String args[])
{
Main abs=new Main();
abs.fun();
}
private void fun()
{
System.out.println("Main function");
}
}
```

## VB .NET

- Visual studio is used to compile and execute vb.net programs
- Select new project then select console applications under visual basic, finally select ok
- Compilation done by visual studio automatically
- Execution done by pressing F5 key

✓ **Criteria : Object oriented based abstraction**

```
Imports System.Console
Module Module1
    Sub Main()
        Dim abs As New child()
        System.Console.WriteLine("I'm in main function")
        WriteLine(abs.fun())
    End Sub
End Module
Public Class parent
    Public i As Integer = 10
End Class
Public Class child
    Inherits parent

Public Function fun() As Integer
        System.Console.WriteLine("I'm in child class")
        System.Console.WriteLine("base class member i values is" & i)
        Return 0
    End Function
End Class
```

Output:
I'm in main function
I'm in child class
base class member i values is10